\documentclass[journal]{IEEEtran}

\usepackage[font=scriptsize]{caption}
\usepackage{vcell}
\usepackage{rotating}
\usepackage{xcolor,soul,framed} 

\colorlet{shadecolor}{yellow}
\graphicspath{{../pdf/}{../jpeg/}}
\DeclareGraphicsExtensions{.pdf,.jpeg,.png}
\usepackage{colortbl}
\usepackage{multirow}
\usepackage{makecell}
\usepackage{hhline}
\usepackage{graphicx}
\usepackage[cmex10]{amsmath}
\usepackage{array}
\usepackage{mdwmath}
\usepackage{mdwtab}
\usepackage{eqparbox}
\usepackage{url}
\usepackage{amssymb}
\usepackage{caption}
\usepackage{subfigure}
\usepackage{siunitx}
\usepackage{gensymb}
\usepackage{float}

\begin{document}
\bstctlcite{IEEEexample:BSTcontrol}
    \title{Stereo X-ray Tomography on Deformed Object Tracking}
  \author{Zhenduo~Shang and
      Thomas~Blumensath

  \thanks{The authors are with the Institute of Sound and Vibration Research at the University of
Southampton, Southampton, United Kingdom.}} 

\markboth{
}{Shang and Blumensath : XXXXX}

\maketitle

\begin{abstract}
X-ray computed tomography is a powerful tool for volumetric imaging, but requires the collection of a large number of low-noise projection images, which is often too time consuming, limiting its applicability. In our previous work \cite{shang2023stereo}, we proposed a stereo X-ray tomography system to map the 3D position of fiducial markers using only two projections of a static volume. In dynamic imaging settings, where objects undergo deformations during imaging, this static method can be extended by utilizing additional temporal information. We thus extend the method to track the deformation of fiducial markers in 3D space, where we use knowledge of the initial object shape as prior information, improving the prediction of the evolution of its deformed state over time. In particular, knowledge of the initial object's stereo projections is shown to improve the method's robustness to noise when detecting fiducial marker locations in the projections of the deformed objects. Furthermore, after feature detection, by using the features' initial 3D position information in the undeformed object, we can also demonstrate improvements in the 3D mapping of the deformed features. Using a range of deformed 3D objects, this new approach is shown to be able to track fiducial markers in noisy stereo tomography images with subpixel accuracy.

\end{abstract}

\begin{IEEEkeywords}
feature detection, X-ray Computed Tomography, image registration, 3D mapping.
\end{IEEEkeywords}

\IEEEpeerreviewmaketitle


\section{Introduction}
\IEEEPARstart{W}{hilst} full X-ray Computed Tomography (XCT) is a mature and widely used volumetric imaging technique applied in various fields, it is relatively slow and individual scans can often take from minutes to several hours. This makes standard XCT unsuitable for imaging of those dynamic processes where objects deform on sub-minute, or even sub-second timescales. One way to overcome this is to only to scan a subset of the full tomographic dataset and then use advanced image reconstruction methods, such as those that utilize Total Variation (TV) constraints or machine learning-based reconstruction, to achieve full reconstructions \cite{Lietal, lee2018deep, ernst2021sinogram, adler2018learned, li20213}. These methods can reduce the number of measurements required to some extent; however, the quality of the reconstructed image is often proportional to the number of measurements acquired. Moreover, these approaches still require considerable time to collect the data, rendering them impractical for fast dynamic imaging. 

Our previous work \cite{shang2023stereo} proposes a stereo X-ray tomography system to recover the 3D locations of simple features such as points and lines using only two stereo projection images rather than reconstructing full tomographic images from limited observations, which typically require strong prior knowledge. Our method avoids the stringent constraints inherent in limited-measurement tomography while remaining effective for general objects that contain basic fiducial markers. Although effective for static objects with relatively few, clearly visible fiducial markers, this method struggled with tracking and estimating dynamically deforming structures in settings where there were too many fiducial markers or where fiducial markers had relatively poor contrast. By treating the dynamic deformation process as a time series of consecutive frames, additional information from consecutive time points can be considered to track fiducials during the deformation process. Similar ideas have previously been explored for the reconstruction from limited projections \cite{zhang2017reducing, zhang2015preliminary, ren2014limited}, whilst \cite{nie2020super} considered using the previous state as the prior information for full reconstruction. However, the quality of the reconstruction depends on the number of scan angles, and so overall scan time and reconstruction quality have to be balanced. Our work extends these ideas to the stereo tomography setting by using ideas from an end-to-end unsupervised model for the 2D/3D image registration, VoxelMorph \cite{balakrishnan2019voxelmorph}. VoxelMorph is a learned model that allows us to incorporate image deformation into a larger, trainable neural network model that can be used to estimate deformations in the 2D projection images as well as the 3D fiducial marker location. 

Similar to our previous stereo tomography work, we are here not seeking full image reconstructions, but are only interested in tracking of fiducial marker locations. We furthermore extend our previous approach by incorporating knowledge of the initial object's structure as prior information within the stereo framework. This enables us to predict and estimate the deformation process over time, achieving both accuracy and temporal consistency. 

\subsection{The stereo X-ray tomography framework}
\begin{figure}[h!]
   \begin{center}
   \includegraphics[width=3.5in]{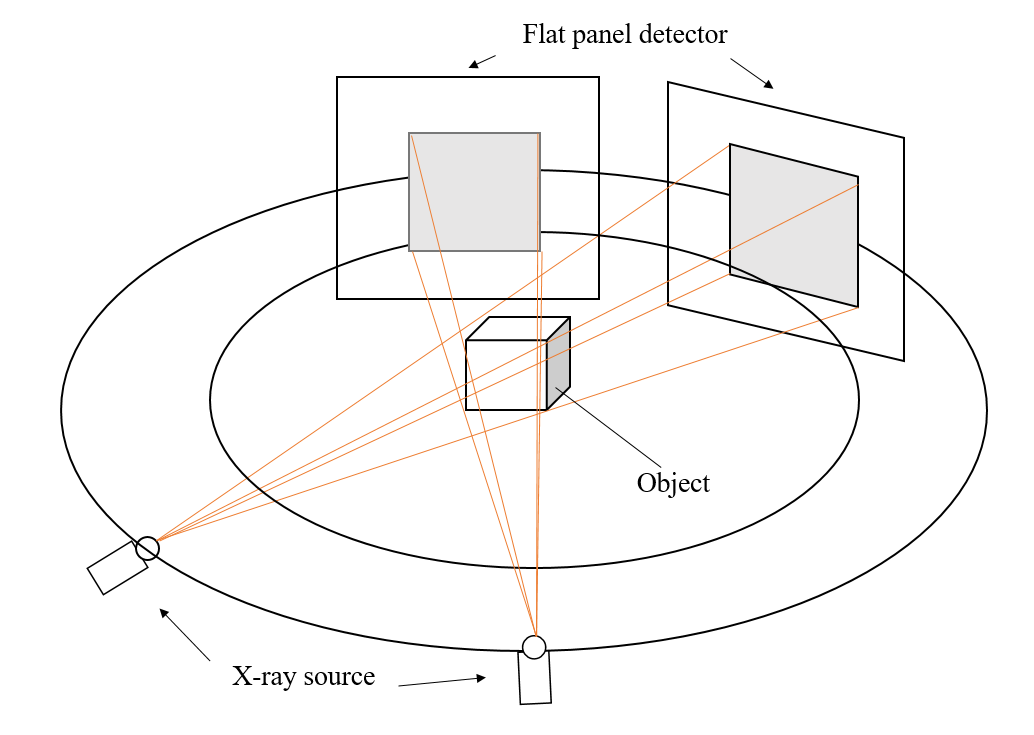}\\
   \caption{For stereo X-ray tomographic imaging with two views, two X-ray projection images are taken of an object from two different viewing directions.}\label{Two X-ray Sources Schematic}
   \end{center}
 \end{figure}

Details of our previous approach can be found here \cite{shang2023stereo}. For completeness, our method is based on a stereo X-ray tomography system setup with two X-ray sources and two detectors, taking images with sub-second temporal resolution, as shown in Fig. \ref{Two X-ray Sources Schematic}. Such a setup would allow us to take stereo projection images of an object at a speed defined by the X-ray flux of the used X-ray source and the readout time of the detector. Crucially, we do not assume that the object (or the source detector pairs) are rotated during data acquisition.


\begin{figure*}[t!]
  \begin{center}
  \includegraphics[width=7in]{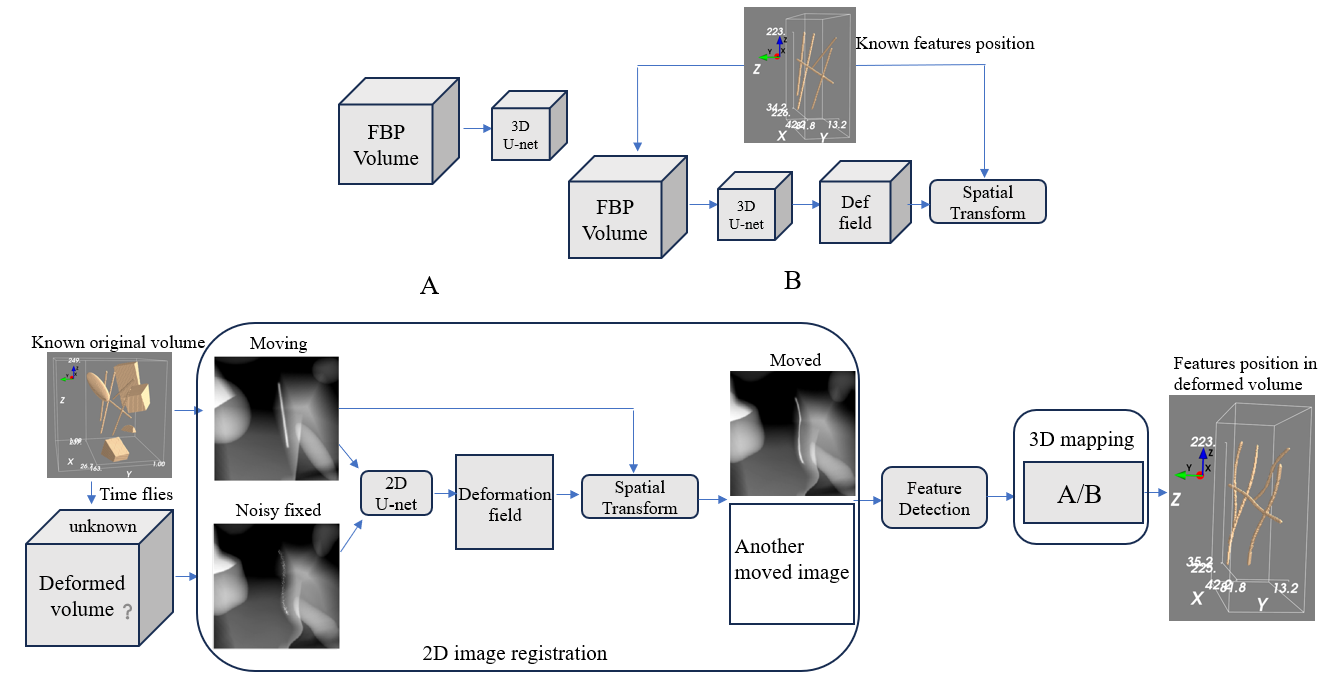}\\
  \caption{The framework of our stereo X-ray tomography system to track deformed line markers. Starting from two frames, a known original volume and its next frame during a deforming process, we denote the projections from the known volume as moving images, and from the deformed volume as the noisy fixed image. The noise during the deforming process makes the projections from the deformed volume difficult to estimate. To locate the markers in the noisy images, we employ the 2D image registration model to estimate the deformation field between the moving and noisy fixed image, warping the deformation to the moving image to obtain the moved image, which is a clean estimate of the noisy fixed image. After the 2D image registration step, our previous stereo X-ray tomography system can be used to extract fiducial markers. After marker identification, the 3D mapping step of our previous approach can be used; however, if there are too many markers, then the original 3D mapping method (A) fails. To overcome this, the features' position in the original known volume is employed as prior information also in the 3D mapping step using another 3D image registration model (B).}\label{overview of framework}
  \end{center}
\end{figure*}

Our previous work developed an algorithm to estimate the spatial location of linear or point fiducial markers within the object or to detect and map point and line like features (such as sharp object edges and corners) \cite{shang2025imagingedgemappingobject}.

Our previous work considered a static setting and so worked with a single pair of X-ray projection images. The method then used a two-stage process. In step one, a 2D U-net is used to identify the location of fiducial markers or low-dimensional object features within the 2D projection images. In a second step, the identified 2D feature maps are back-projected into 3D space using the FDK algorithm and further processed to estimate the exact spatial location within 3D space.

\subsection{Our new approach}

Two challenges exist when using the stereo tomography approach. On the one hand, when imaging fast dynamic processes, limits of X-ray flux can lead to projection images that contain significant amounts of noise, making it difficult to detect the fiducial markers or feature locations accurately. Furthermore, as feature matching between images is not unique, if there are a large number of features that are to be mapped into 3D space, then the 3D mapping algorithm is increasingly likely to make matching errors, which will place features at the incorrect spatial location.

To overcome these issues, the dynamic imaging setting allows us to bring additional information into use. Assuming features move relatively little between individual time steps, their previous location (both in 2D and 3D space) provides valuable additional information that can be used to estimate feature location in 2D and 3D space accurately.

To extend the previous framework to the dynamic setting, we develop a modified approach to the previously used feature detection algorithm as well as an advanced method for 3D mapping. The new framework is shown in Fig. \ref{overview of framework}, consisting of 2D image registration and two alternative 3D mapping solutions. 

Our method thus modifies the previous approach by adding a 2D image registration part to the feature detection network. We here incorporate the 2D image registration model (consisting of a 2D U-net \cite{ronneberger2015u} with a differentiable STN layer \cite{jaderberg2015spatial}) into our estimation model. This network takes both the undeformed `moving image' as well as the noisy `fixed image' to estimate a 2D deformation field. This deformation field can then be applied to the `moving image' to produce a low noise version of the `fixed image'. The low noise image can then be used to estimate feature locations.

We here denote the moving image as $m$, the noisy fixed image as $f_{noisy}$, the moved image as $f_{moved}$ and the spatial transformation network as $STN\left\{ .\right\}$. $U_{2D\alpha}$ is a 2D U-net with its trainable parameters $\alpha$, and the training  loop can thus be expressed as Eq. \ref{eq:1.1}:  
\begin{equation}
    f_{moved}=STN\left\{ m,U_{2D}(f_{noisy},m,\alpha)\right\}\label{eq:1.1}
\end{equation} 

Mirroring our previous work, the feature detection is formulated as a binary classification problem, by training a 2D U-net with parameter $\beta$ to detect the features from the `moved images', expressed as Eq. \ref{eq:1.2}:
\begin{equation}
    f_{feature}=U_{2D}(f_{moved},\beta)\label{eq:1.2}
\end{equation} 

In our original work, once features are detected, the back-projected volume is generated from the two projected feature maps using the FDK algorithm \cite{feldkamp1984practical} and a 3D U-net to estimate features' position in 3D space. We call this method (A). It uses no prior information to estimate features' position in 3D space, which can be formulated as Eq. \ref{eq:1.3} from our previous work. The trainable 3D U-net has parameters $\gamma$ and is trained as a classification network. $V_{feature}$ denotes the features' position in 3D space, the back-projected volume is denoted as $V_{bp}$.
\begin{equation}
    V_{feature}=U_{3D}(V_{bp},\gamma)\label{eq:1.3}
\end{equation} 

However, as method (A) only works if we have very few features, we also utilise the additional information available in our dynamic imaging setting. For the 3D mapping method (B), a 3D image registration model is employed, consisting of a 3D U-net \cite{cciccek20163d} with a differentiable STN layer. This model treats the line features' position in the previous state as the prior information, which is then wrapped with the estimated deformation field to obtain the deformed features' position in 3D space. The loop can be expressed as Eq. \ref{eq:1.4}, where $V_{preFeatures}$ is the line features' position in the previous state, and $\eta$ is the parameter of the 3D U-net.
\begin{equation}
    V_{feature}=STN\left\{ V_{preFeatures},U_{3D}(V_{bp},V_{preFeatures},\eta)\right\}\label{eq:1.4}
\end{equation}

In addition, before 2D image registration, we try to apply our feature detection model on noisy fixed images, to prove that the previous method cannot deal with that why we have to propose a new method.




\section{Dataset}
To train and test our new method, we generate a simulated dataset. We generate 1200 3D images with $256\times256\times256$ voxels each. Each image contains several fiducial line makers and 10 random ellipses. To simulate smooth and continuous deformations, we generated deformation fields by Gaussian filtering (std of 6 pixels, zero mean, unit variance) random matrices. These deformations were applied consecutively to the starting volumes with the ellipses, whilst a different deformation consisting of smooth trigonometric distortions was applied to the lines, keeping the two endpoints fixed. The magnitude of the lines' deformation ranged from $3$ to $6$ pixels. 

The starting volumes and their corresponding deformed volumes are projected at arbitrary angles, generating projections of size $256\times256$ pixels, with a distance between the source and object being 128 voxels in length and a distance between the object and the detector being 128 voxels, i.e. a magnification of 2. To simulate noise, Poisson noise is added to each deformed projection (by plus numpy.random.poisson(img * scale) / scale). We denote these images as `noisy fixed images'. Each data point thus includes the original low noise volume (scale = 10, negligible noise), the projections of the original low noise volume (the so-called moving image) and the noisy fixed projection images (scale = 0.24). As we train the model 2D model on individual pairs of moving and fixed images, taken from the same direction, we generate the training samples with arbitrary projection angles, generating 3600 samples. For each training sample, we keep a copy of the fixed images before adding the noise to provide ground truth data for our deformation estimation method. In this way, we use 1200 pairs of test samples (at $-30\degree$ and $30\degree$) to generate moved images by the trained model, and then these 1200 pairs of moved images are fed into the feature detection model with the well-detected features output, generating 1200 back-projected volumes for training of the 3D mapping step.

There are two methods (A and B) for the 3D mapping step. In method A, 1200 back-projections are used, and their line features' position in the deformed volume is used as ground truth. Method A is the same as the 3D mapping method of the previous work. As for method B, to deal with larger numbers of features, the line features' position in the starting volumes is employed as additional prior information. Thus, one input sample is made by a line features' position from the starting volume and a corresponding back-projected volume, the line features' position in the deformed volume is the ground truth. 



\section{Experimental evaluation}
\subsection{2D image registration}
\begin{figure}[H]
  \begin{center}
  \includegraphics[width=3.5in]{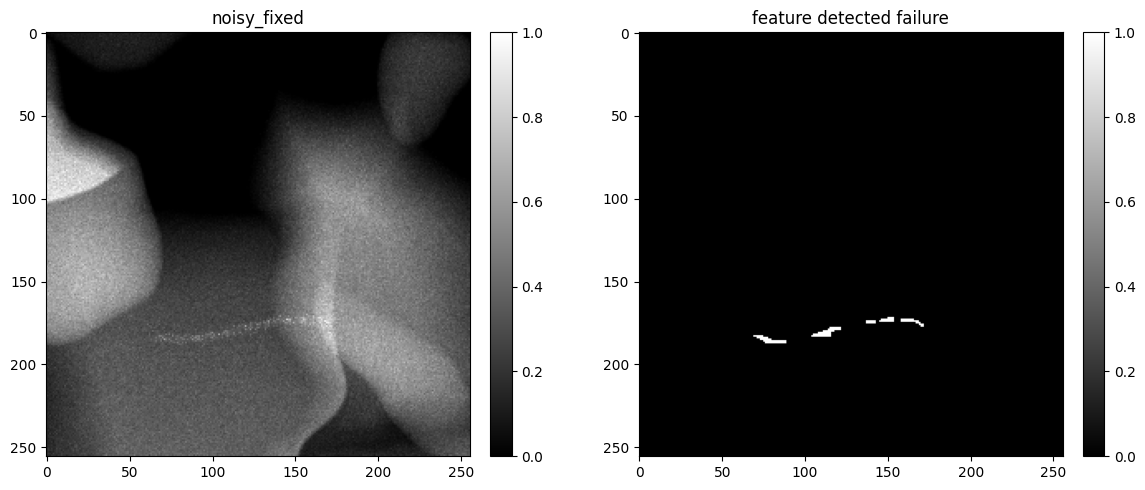}\\
  \caption{The failure of our feature detection method on a noisy projection image. The detection is not clear enough for generating the back-projected volume. Thus, our stereo X-ray tomography is unable to apply directly to this kind of situation; a further processing step is required, and some prior information is employed to improve feature detection.}\label{feature detection failure}
  \end{center}
\end{figure}

To demonstrate the capability of our new framework, we here assume that features, when measured during the deformation process, are accompanied by significant amounts of noise, which severely impacts the quality of the images and prevents accurate feature detection. This causes our previous framework to fail, which is shown in Fig. \ref{feature detection failure}.

\begin{figure}[H]
  \begin{center}
  \includegraphics[width=3.5in]{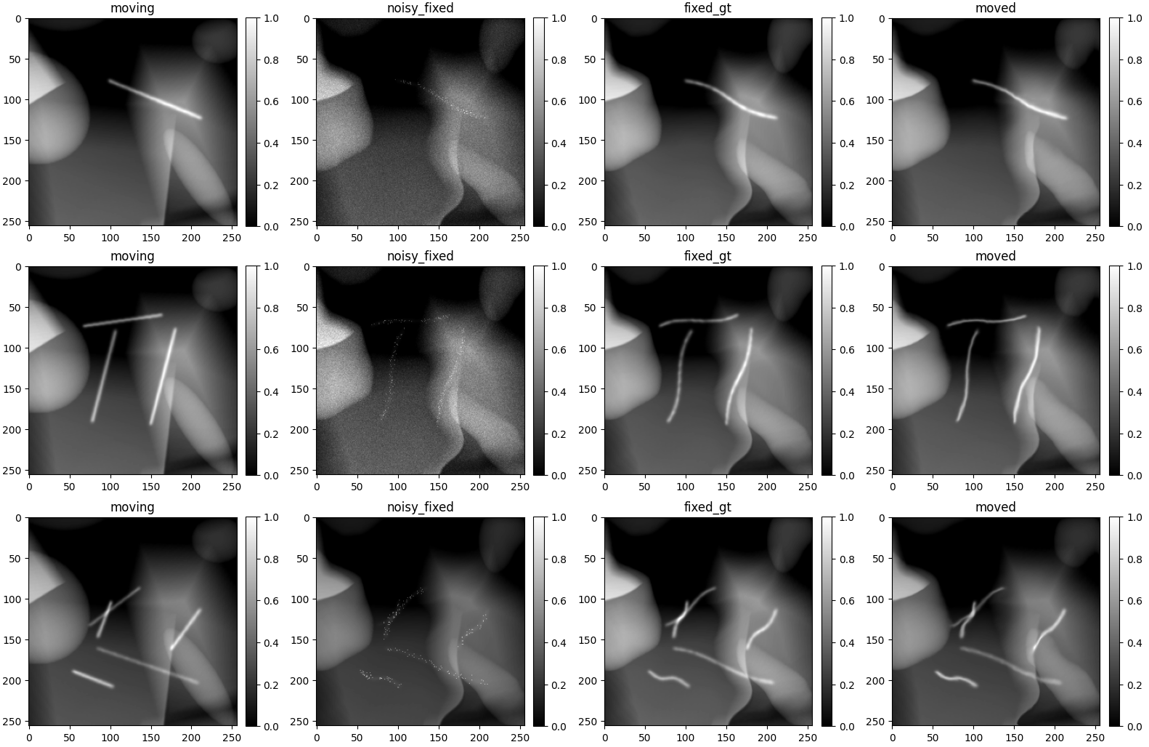}\\
  \caption{2D registration results for data with 1, 3 and 5 lines. The first column is the moving image, the second column is the noisy fixed image, with clear features and background referred to as ground truth, in the third column, and the last column is the estimation, the moved image.}\label{2D registration results}
  \end{center}
\end{figure}

\begin{figure}[H]
  \begin{center}
  \includegraphics[width=3.5in]{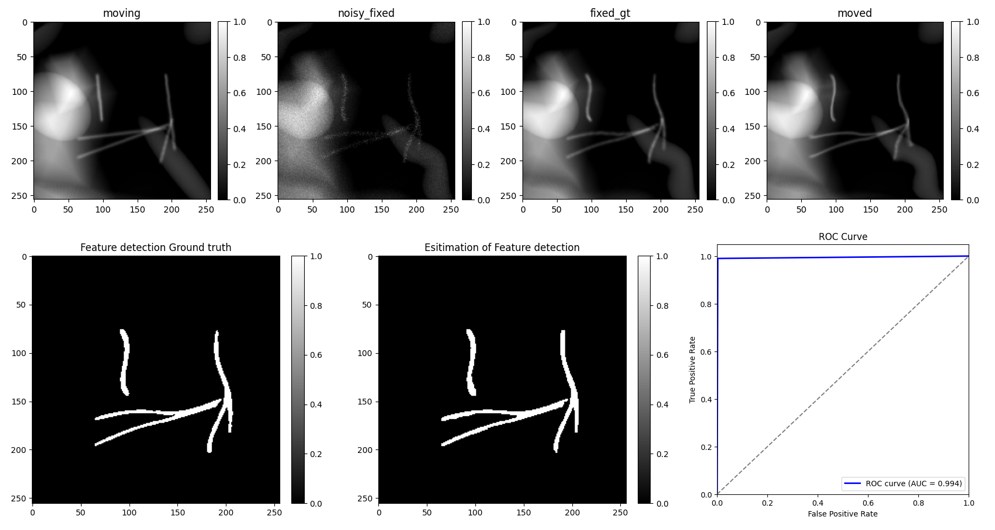}\\
  \caption{A set of test samples on 2D image registration with 5 lines. The top row is the moving image, the noisy fixed image, the ground truth and the estimation. The bottom row shows the estimation of the feature detection, its ground truth, and the ROC curve between them.}\label{2D registration eva}
  \end{center}
\end{figure}

We thus use our new approach to estimate a cleaner fixed image. We present three sets of results with 1, 3 and 5 lines visible in the projections in Fig. \ref{2D registration results}.  The method generally performs well, especially when the lines are not close to each other. When the lines are close, the accuracy decreases clearly. Thus, it's safe to say that the fewer features we have, the better the accuracy will be on the moved images, and it has less chance of having the lines close to each other. Therefore, when our 2D image registration method breaks down, it does not depend on how many features we have; it depends on the chance of features being close and squeezed with each other. To better numerically quantify the accuracy, we operate the feature detection on the moved images first, and then calculate the accuracy of the moved images. A set of test samples with 5 line features is shown as in Fig. \ref{2D registration eva}, the ROC curve shows that the detection has great accuracy, which can indicate the moved image has a good quality from the 2D image registration step. However, a good ROC curve cannot prove that there is no error, and by measuring the distance between lines, we find that an error of between 0 to 3 pixels exists. By considering the radius of the lines, which can be magnified on the detector, this error is of a similar order to the width of the detected lines.

\subsection{3D mapping}
\begin{figure}[H]
  \begin{center}
  \includegraphics[width=3.5in]{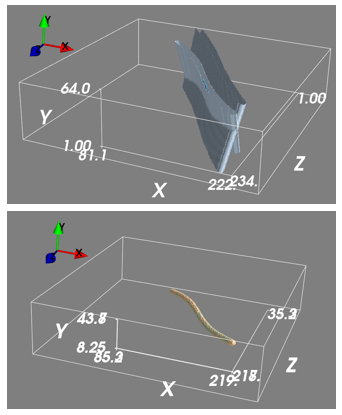}\\
  \caption{A set of samples of single-line cases. The top image is the back-projected volume, and the green and orange colours at the bottom represent the estimation and ground truth.}\label{1 line 3D mapping}
  \end{center}
\end{figure}

\begin{figure}[H]
  \begin{center}
  \includegraphics[width=3.5in]{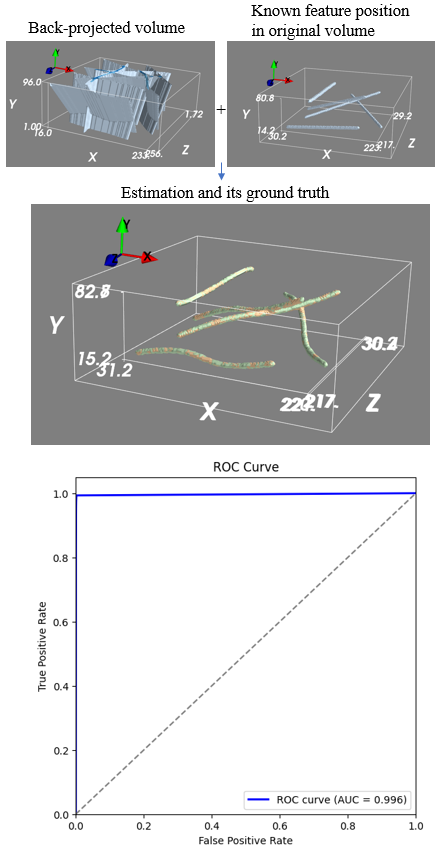}\\
  \caption{A set of samples of a 5-line case and its evaluation of the ROC curve. The top row is the input of our advanced 3D mapping (B) model, and the second row is the output and its ground truth. The last row is the ROC curve between the estimation and the ground truth.}\label{5 lines 3D mapping}
  \end{center}
\end{figure}

Once good feature location maps have been estimated using the 2D model, we then map the features' position into 3D space. When the volume has only 1 feature, our original 3D mapping step still works well, as shown in Fig. \ref{1 line 3D mapping}, which visually achieves a good accuracy. However, our 3D mapping method (A) deteriorates with multiple features in close proximity. We thus evaluate our advanced method in Fig. \ref{5 lines 3D mapping}, which presents a set of results with 5 features. In the rendering of the 3D mapping results, the orange part represents the ground truth, and the green part represents the estimation. As for the evaluation of the 3D mapping, we follow the evaluation process in the following order: first, we ensure that the positions visually overlap well. Based on this, we further conduct a numerical evaluation using the ROC curve and Chamfer distance between the ground truth and the estimation. As shown in the bottom row of the Fig. \ref{5 lines 3D mapping}, the ROC curve shows a good feature overlap, and Chamfer distance \cite{liu2010fast} is used here to further check the quality of the estimation, the value is 0.23 voxels, which indicates a good performance and matches with our visual check. 

\subsection{A demonstration on consecutively deformed samples}
\begin{figure*}[t]
  \begin{center}
  \includegraphics[width=6in]{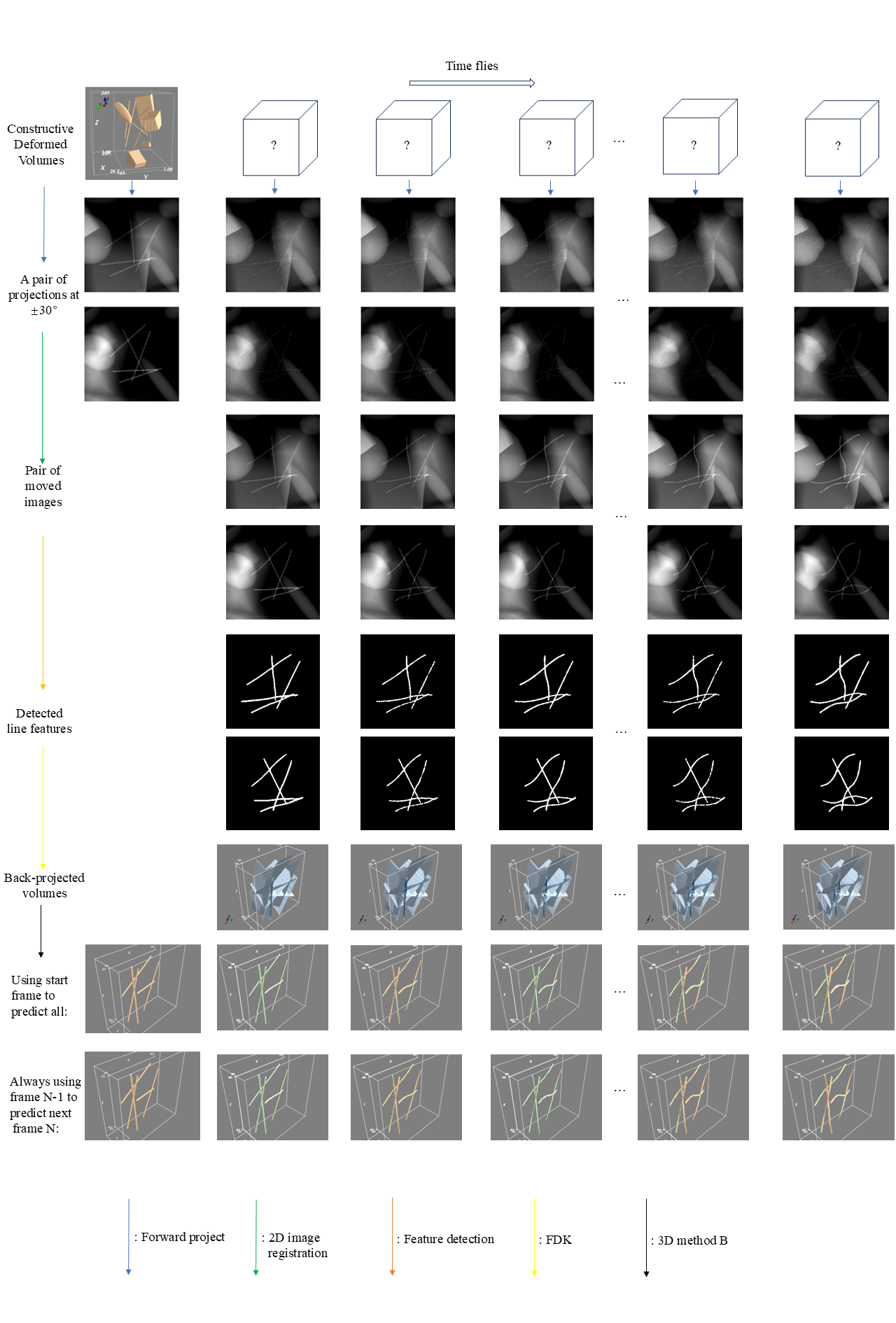}
  \caption{The result of stereo X-ray tomography tracking the deformed features over time. The $1st$ row shows a starting frame from a known object, referred to as the starting frame, and how the inside changed over time is unknown. The $2nd$ and $3rd$ rows showed a pair of noisy projections at $±30\degree$ forward projected by their corresponding frames, where the noise is taken during the deformation. With the help of the projections from the starting frame, the 2D image registration method made the noisy image clear, as shown in the next $4th$ and $5th$ rows: a pair of moved images, which are ready for feature detection operation. With the clear moved images, our feature detection model is employed to detect the line features, shown in the $6th$ and $7th$ rows. In the $8th$ row, the well-detected features generated their back-projected volumes. And in the $9th$ and $10th$ rows, two strategies of the 3D method B showed their estimation. The last row shows the meaning of the arrows with different colours. }\label{constructive frames test}
  \end{center}
\end{figure*}

\begin{figure*}[t]
  \begin{center}
  \includegraphics[width=6in]{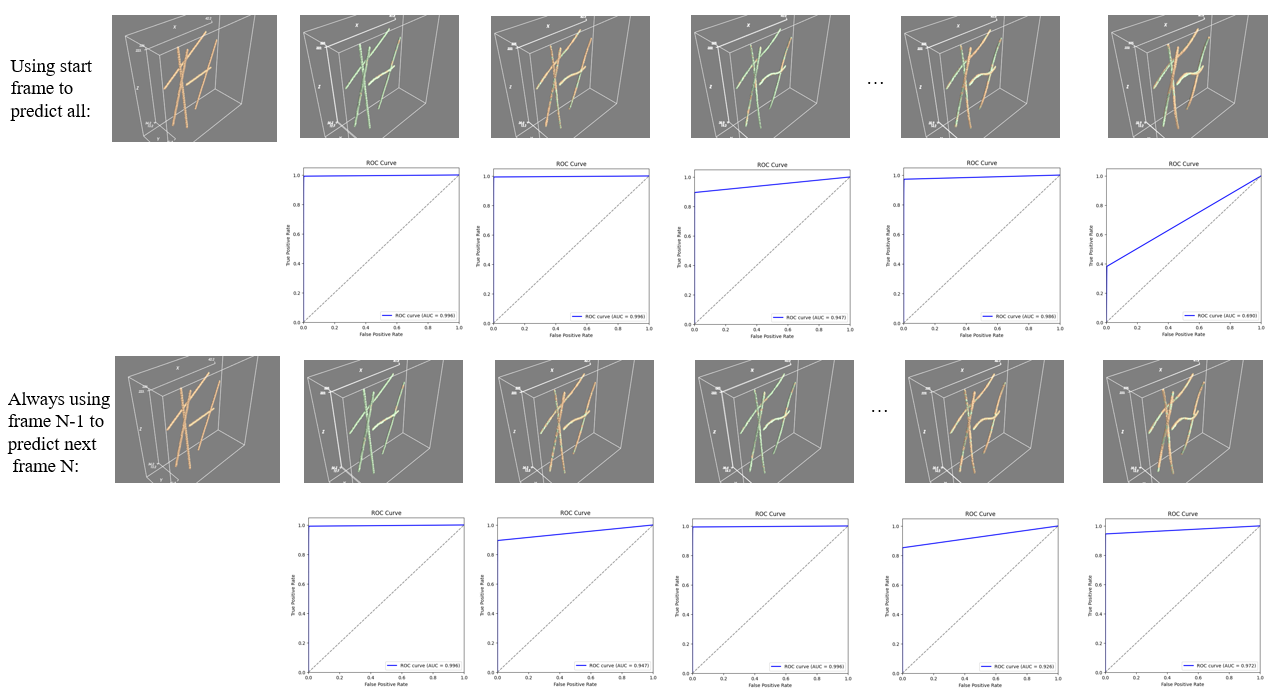}
  \caption{The ROC curve for each frame estimation of two different strategies.}\label{3droc}
  \end{center}
\end{figure*}

To further stretch the capabilities of our approach, we now extend the experiments to a setting where we track features over several frames that change over time. We present a demonstration of the full steps of tracking the features from a consecutively deformed object, shown in Fig. \ref{constructive frames test}. In these results, we do not repeatedly focus on the performance of 2D image registration and 3D mapping, as discussed in detail in the above section. We instead focus on the performance of the two different strategies after the 3D mapping using method B. The first 3D mapping here uses the positions of features in the starting frame to predict those of frame 2, frame 3, ..., N by training a general model with a proper deformation range, which is as big as possible. However, the limitation of this strategy is that if the deformation on frame N is too big for the model to predict, it will be. Thus, training the model with a slightly different dataset, using frame (N-1) as the prior information to predict frame N (for example: using the start frame to predict frame 2, then using the estimated frame 2 to predict frame 3, and so on), gives our method better generalization ability. Fig. \ref{constructive frames test} shows the comparison of these two strategies. In the test results of using a start frame to predict all, the predicted frames 1, 2 and 3 have an accuracy of less than 1 voxel. A visual error arises when predicting frame N-1, and an even clearer error happens when predicting frame N. As for the strategy of using frame N-1 to predict frame N, all the estimations have a good accuracy, all with less than 1 voxel error. This approach thus has a clear advantage compared with the former strategy from the same test samples, showing a better generalisation ability. To further quantify the accuracy, we use the ROC curves, as shown in Fig. \ref{3droc}. In the estimation of frame N by the first strategy, the ROC curve showed a sharp drop in performance, which matches the visual errors. The ROC curve from the second strategy always shows good performance.

The error evaluated by the Charmfer distance again demonstrates similar results. The estimation of frame N using the start frame to predict all other frames sharply increases the distance to $2.40$, much higher than the rest of the other errors, which are all smaller than $0.40$. By using frame N-1 to predict the next frame N, the Chamfer distances are all smaller than $0.34$. 

\section{Discussion and conclusions} 
In this paper, we extend our initial stereo X-ray tomography system from static to dynamic, from a general framework of extracting the features' position in 3D space with only two projections to tracking the features' position in 3D space during a deforming process over time, and from applying to very few basic fiducial markers to relatively many basic fiducial markers. In the test of 2D registration with 1, 3 and 5 fiducial line markers, the error stays between 0 to 3 pixels in terms of fiducial location, which is a relatively good performance considering the radius of the fiducial line markers projected onto the detectors. In testing the 3D mapping, two strategies are tested on a series of consecutive frames of 5 lines deforming over time. Both strategies have a good accuracy within 1 voxel error at their suitable deformation range, but the latter method shows a better generalisation ability. Whilst we have here explored the method using simulated data, testing with real-world samples will remain for a future study.

\section{Acknowledgment} 
 The authors would like to thank the University of Southampton for the use of the IRIDIS High Performance Computing Facility and associated support services.
\ifCLASSOPTIONcaptionsoff
  \newpage
\fi





\bibliographystyle{IEEEtran}
\bibliography{IEEEabrv,Bibliography}

\vfill


\end{document}